\documentclass{webofc}
\usepackage[varg]{txfonts}   

\usepackage{subfig, ulem}
\usepackage{multirow}

\woctitle{New Frontiers in Physics 2014}
\begin{document}
\title{The CUORE and CUORE-0 Experiments at Gran Sasso}
%
%

\author{A.~Giachero         \inst{1,2} \fnsep\thanks{\email{Andrea.Giachero@mib.infn.it}} \and
        D.~R.~Artusa        \inst{3}     \and
        F.~T.~Avignone~III  \inst{3}     \and
        O.~Azzolini         \inst{4}     \and 
        M.~Balata           \inst{5}     \and
        T.~I.~Banks         \inst{6,7}   \and
        G.~Bari             \inst{8}     \and
        J.~Beeman           \inst{9}     \and
        F.~Bellini          \inst{10,11} \and
        A.~Bersani          \inst{13}
% ...
}

\institute{Dipartimento di Fisica, Universit\`a di Milano-Bicocca, Milano I-20126 - Italy                      \and 
           INFN - Sezione di Milano Bicocca, Milano I-20126 - Italy                                            \and 
           Department of Physics and Astronomy, University of South Carolina, Columbia, SC 29208 - USA         \and 
           INFN - Laboratori Nazionali di Legnaro, Legnaro (Padova) I-35020 - Italy                            \and 
           INFN - Laboratori Nazionali del Gran Sasso, Assergi (L'Aquila) I-67010 - Italy                      \and 
           Department of Physics, University of California, Berkeley, CA 94720 - USA                           \and 
           Nuclear Science Division, Lawrence Berkeley National Laboratory, Berkeley, CA 94720 - USA           \and 
           INFN - Sezione di Bologna, Bologna I-40127 - Italy                                                  \and 
           Materials Science Division, Lawrence Berkeley National Laboratory, Berkeley, CA 94720 - USA         \and 
           Dipartimento di Fisica, Sapienza Universit\`a di Roma, Roma I-00185 - Italy                         \and 
           INFN - Sezione di Roma, Roma I-00185 - Italy                                                        \and 
           Dipartimento di Fisica, Universit\`a di Genova, Genova I-16146                                      \and 
           INFN - Sezione di Genova, Genova I-16146 - Italy
% ...
}


\abstract{The Cryogenic Underground Observatory for Rare Events (CUORE) is an experiment to search for neutrinoless double beta decay ($0\nu\beta\beta$) in $^{130}$Te and other rare processes. CUORE is a cryogenic detector composed of 988 TeO$_2$ bolometers for a total mass of about 741~kg. The detector is being constructed at the Laboratori Nazionali del Gran Sasso, Italy, where it will start taking data in 2015. If the target background of 0.01 counts/(keV$\cdot$kg$\cdot$y) will be reached, in five years of data taking CUORE will have an half life sensitivity around $1\times 10^{26}$~y at 90\% C.L. As a first step towards CUORE a smaller experiment CUORE-0, constructed to test and demonstrate the performances expected for CUORE, has been assembled and is running. The detector is a single tower of 52 CUORE-like bolometers that started taking data in spring 2013. The status and perspectives of CUORE will be discussed, and the first CUORE-0 data will be presented.}
\maketitle
\section{Introduction}\label{sec:intro}
The interest in neutrino physics has increased in the recent years since the discovery of neutrino oscillations. Experiments measuring oscillations in solar, atmospheric, accelerator, and reactor neutrinos have made tremendous progress in pinning down the neutrino mixing angles and oscillation frequencies, which enter as parameters in the Standard Model with the inclusion of the nonvanishing neutrino masses. To complete this scenario next-generation experiments are needed to study the mass hierarchy of the neutrino mass eigenstates and the neutrino-antineutrino dichotomy (\textit{Majorana particle}: $\nu\equiv\overline{\nu}$ or \textit{Dirac particle}: $\nu\not\equiv\overline{\nu}$). One practical way to investigate these open issues is to search for neutrinoless double beta decay ($0\nu\beta\beta$)~\cite{Avignone_RevModPhys_80_481}. The observation of this process would prove that neutrinos are Majorana particles, and could provide other fundamental information about the neutrino mass.

The CUORE experiment~\cite{CUORE1,CUORE2,CUORE3}, will search $0\nu\beta\beta$ of the isotope $^{130}$Te with a 741$\,$kg array of cryogenic bolometers. The detector is being built at the Laboratori Nazionali del Gran Sasso (LNGS), Italy, and will start data taking in 2015. In the meanwhile a smaller scale prototype experiment named CUORE-0, with a 39$\,$kg detector mass, has been operating at LNGS since March 2013. The main purpose of CUORE-0 is to check the effectiveness of the material cleaning and detector assembly procedures developed for CUORE, but it is a sensitive $0\nu\beta\beta$ experiment in its own right.

In this report, after a brief introduction to double beta decay, the bolometric technique will be described and the status of the CUORE and CUORE-0 experiments will be presented.

\section{Neutrinoless double beta decay}\label{sec:dbd}
Double beta decay is a rare process in which a nucleus changes its atomic number by two units. It is possible only in nuclei with an even number of neutrons and protons, in which the single beta decay is energetically forbidden. One mode of double beta decay, so-called 2 neutrino more has been experimentally observed for several nuclei; a second mode, so-called neutrinoless mode, is theoretically predicted by theories beyond the Standard Model but has not been observed to date. In two-neutrino double beta decay ($2\nu\beta\beta$) two electrons and two anti-neutrinos are emitted (equation \ref{eq:2nubetabeta}). 

\begin{equation}\label{eq:2nubetabeta}
 N(A, Z)\rightarrow N(A, Z+2)+2e^-+2\overline{\nu}_e\quad(2\nu\beta\beta)
\end{equation}

This decay mode is allowed by the standard model and has been observed in all the double beta decay candidate nuclei that are considered interesting from an experimental perspective. It is in fact the rarest weak decay ever observed, with half lives in the range $(10^{19}\div 10^{21})$ years.

In neutrinoless double beta decay ($0\nu\beta\beta$) only the two electrons are emitted (equation \ref{eq:0nubetabeta}). The decay violates the lepton number conservation by two units and it is possible only if the neutrino is a massive Majorana particle~\cite{Frampton_PhysRevD_25_1982}.

\begin{equation}\label{eq:0nubetabeta}
N(A, Z)\rightarrow N(A, Z+2)+2e^-\quad(0\nu\beta\beta)
\end{equation}

Its transition width is proportional to the square of the effective Majorana mass, $|\langle m_{ee}\rangle|$. In case of neutrino mixing the Majorana mass experiments measure a specific mixture of neutrino mass eigenvalues, $|\langle m_{ee}\rangle|^{2} =|\sum_{i} U_{ei}^{2} m_{i}|^{2}$, \textit{i} summed over all mass eigenstates. From the $0\nu\beta\beta$ half-life it is therefore possible to infer important information concerning the mass hierarchy and the absolute mass scale of neutrinos.

In $0\nu\beta\beta$ the nucleus is heavy enough that all the energy is shared between the two electrons and the recoil is negligible. Thus the experimental signature is a monochromatic line at the $Q$-value of the decay. The experimental sensitivity ($S^{0\nu}$), defined as the decay time corresponding to the minimum number of detectable events above background ($B$), in given by

\begin{equation}\label{eq:sens} 
S^{0\nu}=\ln{2}\cdot\epsilon\cdot\frac{\mathrm{i.a.}}{A}\cdot\sqrt{\frac{M\cdot T}{B \cdot \Gamma}}
\end{equation}
where i.a. is the isotopic abundance of the a chosen $\beta\beta$-emitter isotope, $M$ its mass, $A$ its molecular mass, $\epsilon$ the efficiency of the detector, $\Gamma$ the energy resolution (around the $Q$-value), $B$ is the background index, expressed in counts/(keV$\cdot$kg$\cdot$y), and finally, $T$ the measurement time. According to equation \ref{eq:sens}, it is straightforward that to obtain the best sensitivity a double beta decay experiment must have a very large mass, high efficiency, a good resolution, a long measurement time, a very low background and the chosen $\beta\beta$-emitter isotope should have a high natural isotopic abundance. If enrichment is not necessary the result is a significant cost savings. Although all the above parameters must be optimized for an experiment to be competitive, the reduction of the background represents the most challenging aspect.

\section{The CUORE bolometers}\label{sec:bolo}
CUORE (Cryogenic Underground Observatory for Rare Events) will use the bolometric technique to search for $0\nu\beta\beta$ in $^{130}$Te. This technique was proposed by E. Fiorini and T.O. Niinikoski in 1984~\cite{Fiorini} as an alternative to the more conventional enriched $^{76}$Ge diodes.  

\begin{figure}[t]
\begin{centering}
 \begin{tabular}{l c}
 \subfloat[]{\includegraphics[width=0.55\columnwidth]{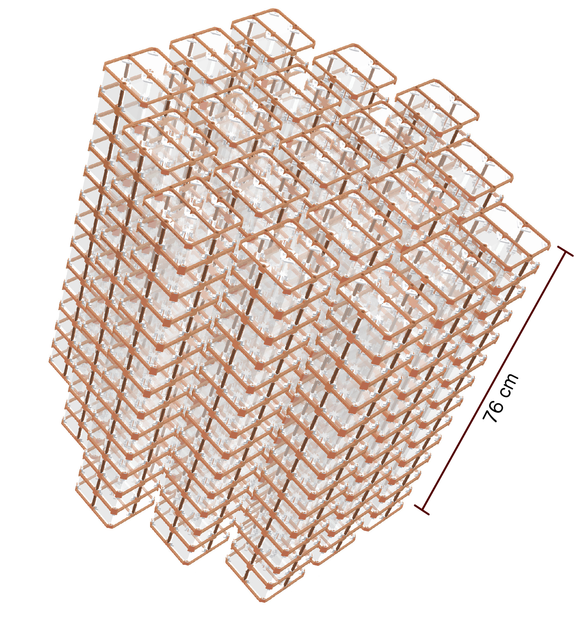}} &
   \hspace{0.02\columnwidth}
 \multirow{-28}[6]{*}{\subfloat[]{\includegraphics[width=0.35\columnwidth]{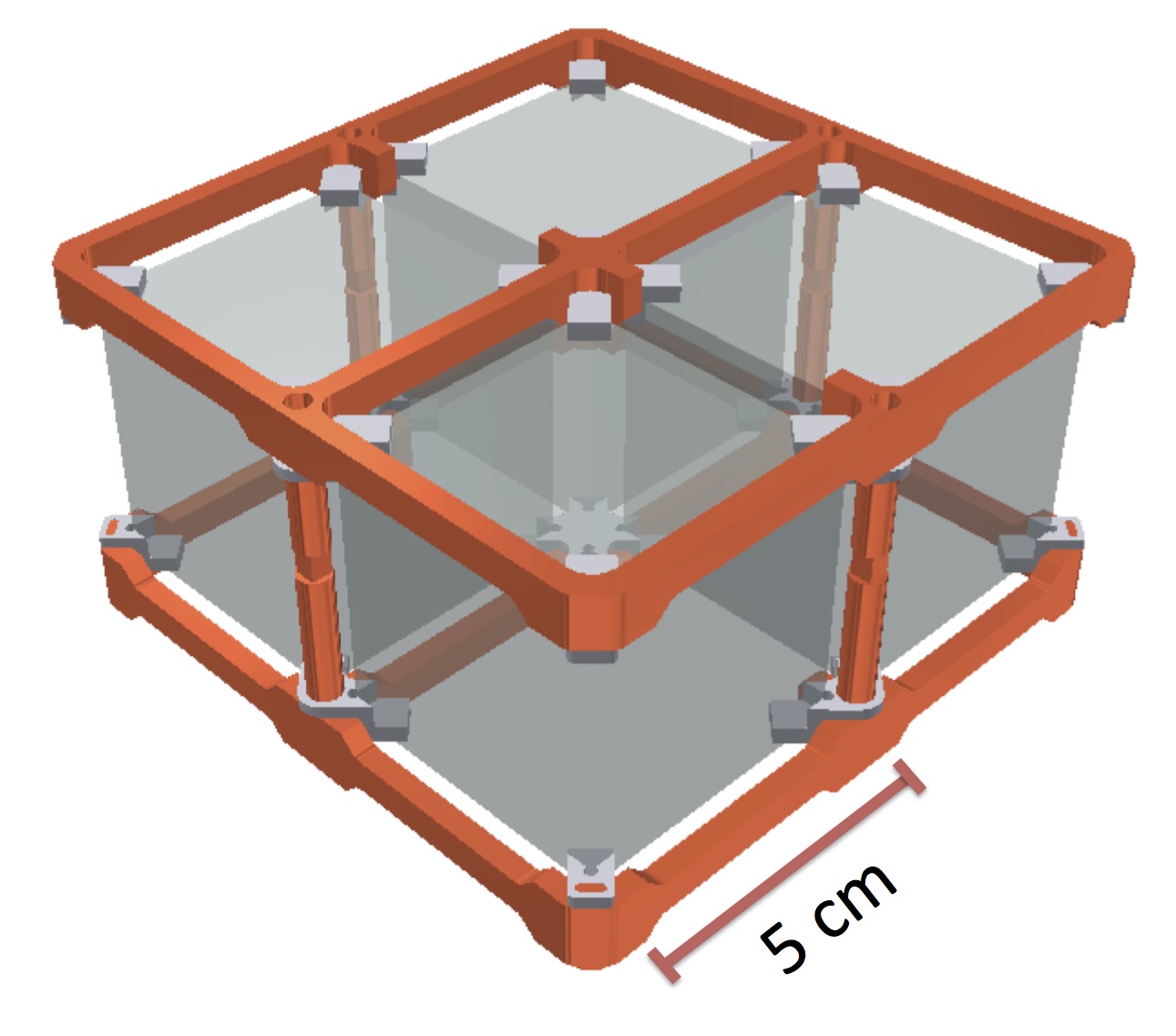}}}\\
 & \multirow{-10}[6]{*}{\subfloat[]{\includegraphics[width=0.35\columnwidth]{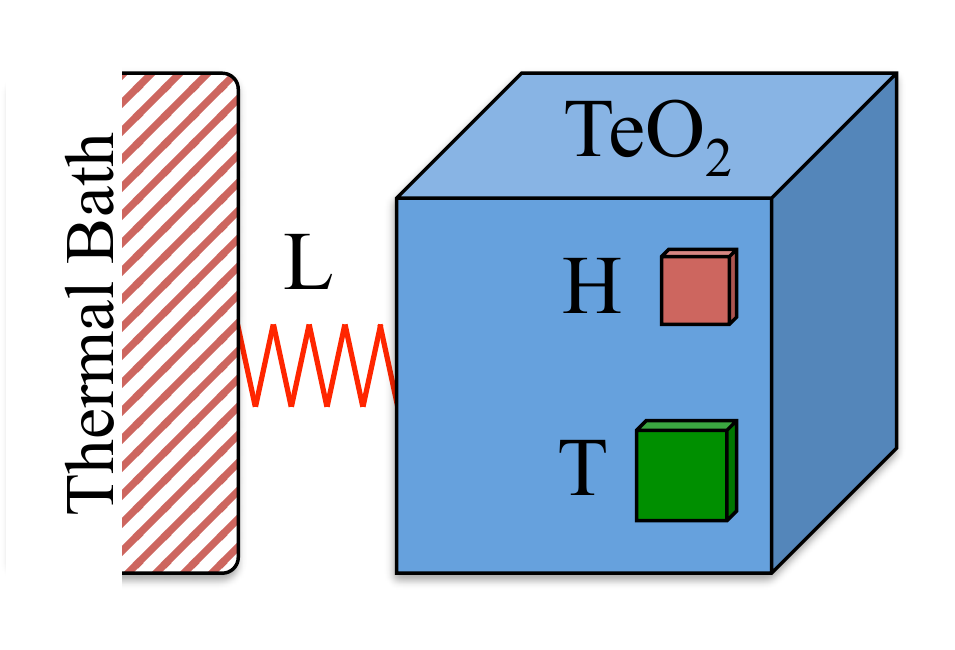}}} \\
 \end{tabular}
\caption{(a)~Illustration of the planned 19-tower CUORE detector array. 
(b)~Closeup of a single tower floor showing four TeO$_2$ crystals held inside their copper frame by PTFE spacers. 
(c)~Schematic diagram of an individual TeO$_2$ crystal bolometer. Each crystal is instrumented with a heater~(H) and a thermistor~(T); the PTFE spacers and sensor readout wires act as weak thermal links~(L) between the crystal and the thermal bath of the copper frame. Figure reprinted from~\cite{CUORE2}. \label{fig:cuorebolo}}
 \end{centering}
\end{figure}

A Bolometer is as a two-component object composed of: an \textit{energy absorber} in which the energy deposited by a particle is converted into phonons, and a \textit{sensor} that converts thermal excitations into a readable electric signal variation (figure \ref{fig:cuorebolo} \textit{(c)}). The absorber must be coupled to a constant temperature bath by means of an appropriate thermal conductance.

The temperature rise $\Delta T$ is related to the energy release $E$ and, denoting by $C$ the heat capacity of the bolometer, can be written as $\Delta T = E / C$. The accumulated heat flows then to the heat bath through the thermal link and the absorber returns to the base temperature with a time constant $\tau = C/G$ where $G$ is the thermal conductance of the link: $\Delta T(t)=E/C\,e^{-t/\tau}$. $\tau$ should be long enough to allow the development of a high resolution signal but short enough to avoid pile-up. In order to obtain a measurable temperature the heat capacity and the base temperature of the absorber crystal must be very small. For these reasons bolometers are usually operated at a temperature around 10~mK and are made of dielectric materials, so that only the lattice heat capacity plays a role. By using large mass bolometers it is possible to achieve excellent energy resolution, comparable to that obtained with germanium detectors. They have also the advantage that they can be built with a wide range of materials, so several isotopes could be studied with this technique. Their main drawback is that the thermal origin of the signal makes them intrinsically slow.

The temperature rise resulting from a single nuclear decay is measured by a Neutron Transmutation Doped (NTD) thermistor. These are glued on the absorber crystal via a dot matrix of bicomponent epoxy. By suitable biasing of the NTD and amplification of the voltage variation across it, thermal pulses are converted to electrical signals compatible with conventional data acquisition system. When operated at $\simeq 10$~mK the CUORE bolometers show an heat capacity of $\simeq 10^{-9}$~J/K, and the NTD sensors have a resistance of few tens of M$\Omega$. An energy release of 1~MeV produces a temperature rise of $\simeq100~\mu$K, which translates to a resistance variation of about 3~M$\Omega$ and a voltage variation of $100~\mu$V at the loads of the sensor. The signal has a typical time evolution of $\simeq5$~s with a typical rise and fall times in the range $t_r = (40 \div 80)~\mathrm{ms}$ and  $t_f = (100 \div 700)~\mathrm{ms}$. The thermal response of a bolometer varies with its temperature, so each crystal is also instrumented with a Si resistor. The signals induced by these Joule heaters are very similar in shape to those produced by particle interactions, and are used in post-processing to correct the detector gain variations induced by the thermal instabilities of the cryogenic system.


\section{The CUORE experiment}\label{sec:exp}
CUORE aims at searching for $0\nu\beta\beta$ of  $^{130}$Te using the bolometric technique. The CUORE detector is an array of 988 cryogenic bolometers arranged in 19 towers of 52 crystals (figure \ref{fig:cuorebolo} \textit{(a)}). Each tower consists of 13 floors of 4 crystals each (figure \ref{fig:cuorebolo} \textit{(b)}). The tower structure is made of ultra pure copper and the crystals are coupled to it by mean of small PTFE supports. The bolometer is a TeO$_2$ cubic crystal (\textit{absorber}), $5\times 5\times 5$~cm$^3$ in size, with an NTD-Ge thermistor (\textit{sensor}) glued onto the crystal surface. Tellurium is an advantage in this instance because of the relatively high natural abundance (34.2\%~\cite{Fehr}) of the $0\nu\beta\beta$ candidate isotope, which means that enrichment is not necessary to achieve a reasonably large active mass. Also, the Q-value (around 2528~keV~\cite{Redshaw,Scielzo,Rahaman}) of the decay falls between the peak and the Compton edge of the 2615~keV gamma line of $^{208}$Tl, the highest-energy gamma from the natural decay chains; this leaves a relatively clean window in which to look for the signal. Moreover crystals or TeO$_2$ can be grown to large size with low radioactive contamination~\cite{Xtals}. They have good mechanical properties and low heat capacity at low temperature (dielectric and diamagnetic). The entire detector will be maintained at the temperature of about 10~mK through a pulse tube assisted cryostat~\cite{CUORE_CRYO1,CUORE_CRYO2}. 

Stringent radiopurity constraints must be imposed on all the materials facing the detectors and on the detectors themselves and all the assembly procedures must be performed underground and must be designed to avoid possible recontamination during the detector construction. The materials radiopurity was verified with low-background measurement techniques. The TeO$_2$ crystals were manufactured by the Shanghai Institute of Ceramics, Chinese Academy of Science (SICCAS), following a strict procedure defined by the CUORE collaboration~\cite{Xtals}. From each batch a few crystals were promptly delivered to LNGS and tested as bolometers to verify their radiopurity and to check their performance as detectors~\cite{CCVR}. Dedicated procedures were developed for the handling and cleaning of each detector component. In particular the cleaning procedure for the copper frames was verified with a dedicated bolometric measurement~\cite{TTT}. To avoid possible recontamination after cleaning, all the detector assembling procedures are performed in glove boxes flushed with nitrogen.
The tower assembling procedure consists essentially of three steps: 1) the NTD-sensor and the Si-heater are glued on the crystal; 2) the crystals are assembled together in a single CUORE tower, and wire trays are mounted on the sides of the tower to bring the signals from the crystals to the mixing chamber of the dilution refrigerator; 3) for each bolometer the sensor and the heater are bonded to the pads present on the wire trays with 25~$\mu$m thick gold wires. To guarantee high success rate and reproducibility, these steps are performed in underground clean room with custom designed tools that make the whole assembly procedure semi-automatic. Completed towers are stored in nitrogen-flushed canisters to await future installation in the cryostat. 

The CUORE experiment is currently in the construction phase at LNGS, the detector installation is scheduled for spring 2015. All 19 towers have been assembled, instrumented, and stored in nitrogen-flushed atmosphere while waiting to be installed in the cryostat. The cryostat was cooled down to 4~K by mean of pulse tubes, and all heat leaks were repaired. The detector calibration system was installed and successfully tested at 4~K. The dilution unit was previously tested in its own custom cryostat, where it reached a base temperature of 5~mK, and subsequently in the CUORE cryostat where it reached a base temperature of 8~mK. Before the end of 2014 the first bolometric tests of the cryostat will be performed. A mini-tower made of CUORE-like crystal will be cooled and signal will be acquired and analyzed. Starting from the middle of 2015, the mounting of the towers on the cryostat will take place and the signal wires will be connected. In parallel, the readout electronics and the data acquisition system will be installed.

\section{The present: CUORE-0}\label{sec:c0}
CUORE-0~\cite{CUORE0} is a single CUORE-like bolometer tower that has been operating at LNGS since March 2013. This pilot experiment has two main goals: the test of all the cleaning and assembling procedures designed for CUORE, and the high statistics test of the improvements achieved in the background reduction with respect to the completed  Cuoricino experiment~\cite{CUORICINO}. The tower is installed in Hall A at LNGS in the former Cuoricino cryostat. It is composed of 52 natural TeO$_2$ bolometers, each 5~cm cube. The total detector mass is 39~kg ($\simeq 11$~kg in $^{130}$Te).

\begin{figure}
\centering
\includegraphics[width=\textwidth]{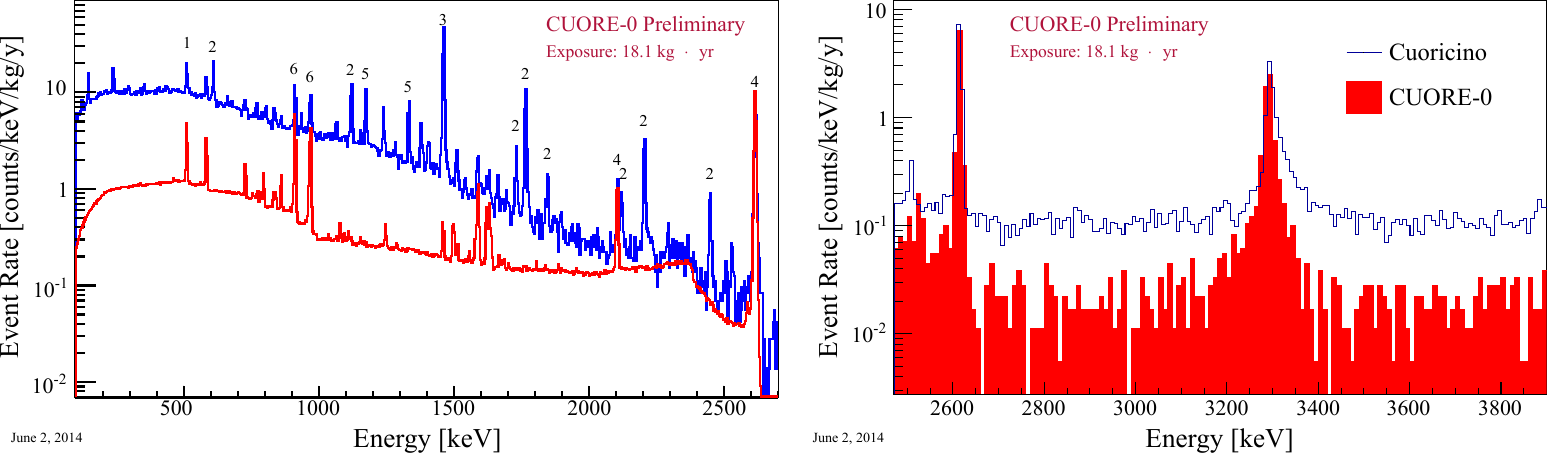}
\caption{Left: (CUORE-0 calibration spectrum (red) and CUORE-0 background spectrum (blue). For convenience, it is normalized to have the same intensity of the 2615 keV line of $^{208}$Tl as measured in the non-calibration spectrum. The labels in the plot correspond to identified radioactive peaks: e$^{+}$e$^{-}$ annihilation (1), $^{214}$Bi (2), $^{40}$K (3), $^{208}$Tl (4), $^{60}$Co (5) and $^{228}$Ac (6). Right: CUORE-0 (red histogram) and Cuoricino (blue histogram) background in the $\alpha$ region. The comparison shows reduction of the flat background caused by degraded $\alpha$ particles in the energy region of $(2.7\div 3.1)$~MeV and $(3.4\div 3.9)$~MeV. The peak at 3.2 MeV is due to $^{190}$Pt contamination in the crystals.}
\label{fig:spectra}       
\end{figure}

The cryogenic apparatus is shielded by 20~cm thick lead shield, to absorb $\gamma$-rays, and by 20~cm of borated polyethylene, to slow down and absorb neutrons. The detector is protected from radioactive contamination in the cryostat materials by 1~cm of low-activity lead in thermal equilibrium with a 600~mk heat shield. The whole system is enclosed in a Faraday cage that suppresses electromagnetic disturbances. The bolometers are calibrated about once per month by inserting a pair of $^{232}$Th source wires between the cryostat walls and the external lead shield. The calibration and background spectra are shown in figure~\ref{fig:spectra} (left). The total exposure acquired between March 2013 and May 2014, and summed over all the detectors, is 18.1$\,$kg$\cdot$y. 

From the study of the dataset of Cuoricino, the previous experiment, it turned out that the background on the $0\nu\beta\beta$ region (ROI: $2470\div 2570$~keV) was composed principally of three components: $\gamma$ events from $^{232}$Th contamination in the cryostat materials (about 30\%), contamination from the crystal surfaces (about 10\%), and from the surfaces of the copper structure (about 50\%). In CUORE-0 the first of these three components it is expected to be comparable with the value extrapolated from Cuoricino, because the two experiments are hosted in the same cryostat. In CUORE this contribution is expected to be negligible thanks to better shielding of the detectors from the cryostat. The other two background components are due to $\alpha$-decays taking place on the surface of the crystals or of the copper structure surrounding them. If the decays occur very close to the surface, the decay products ($\alpha$s and $\beta$s) can escape and reach the active volume of the detector, depositing therefore only a fraction of their energy. This contribution gives a continuous flat spectrum that extends from the Q-value of the decay ($>4$~MeV) down to the energy region where the $0\nu\beta\beta$ signal is expected. This kind of background contribution is expected to be reduced in CUORE-0 with respect to Cuoricino thanks to the improvements in the material cleaning procedures developed for CUORE. A way to quantify this background reduction is to study the energy spectrum above the 2615~keV $^{208}$Tl peak. In fact, above this energy the $\gamma$ background becomes negligible, and the $\alpha$ background dominates.

\begin{figure}
\centering
\includegraphics[width=\textwidth]{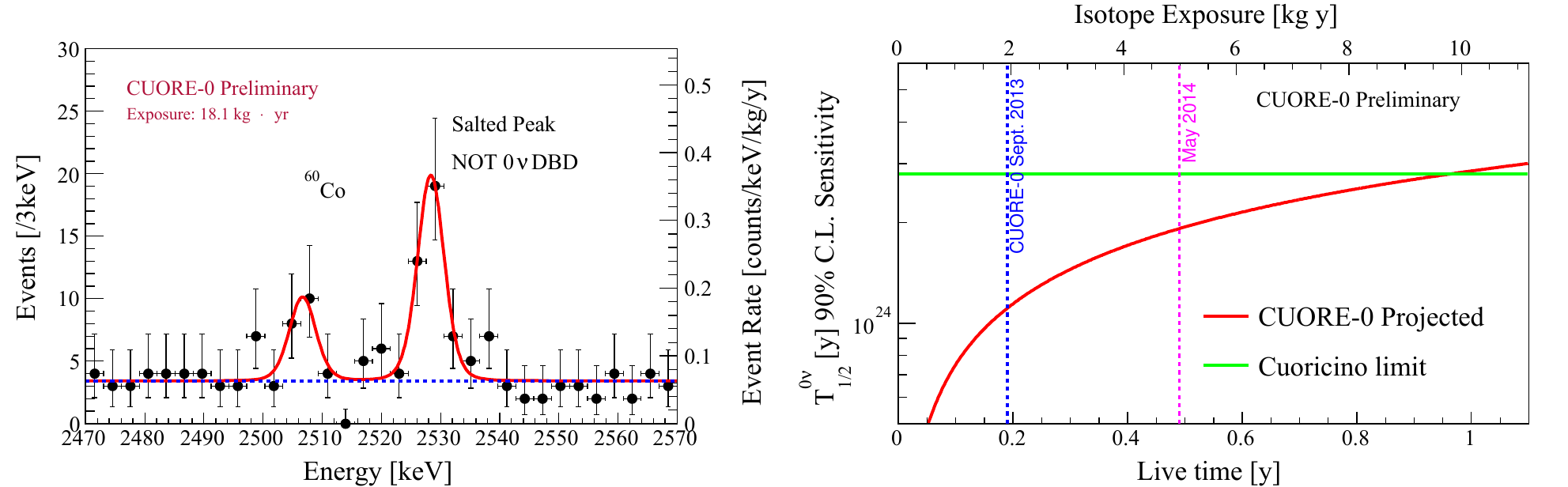}
\caption{Left: $0\nu\beta\beta$ energy region of interest. The peak at 2506~keV is due to the sum of the two $\gamma$s from $^{60}$Co. The peak at 2528 keV is the salted $0\nu\beta\beta$ peak (see text). Right: Projected sensitivity to the $0\nu\beta\beta$ half-life. CUORE-0 sensitivity will overcome the Cuoricino half life limit within about one year of live time.}
\label{fig:salted}       
\end{figure}

For comparison, in figure \ref{fig:spectra} (right), the CUORE-0 and Cuoricino energy spectra in the $\alpha$ region are superimposed. The plots demonstrate the effectiveness of the new surface cleaning procedures developed for CUORE. The flat $\alpha$ background index is (0.20$\pm$0.001)$\,$counts/(keV$\cdot$kg$\cdot$y) in CUORE-0 while it was (0.110$\pm$0.001)$\,$counts/(keV$\cdot$kg$\cdot$y) in Cuoricino, a reduction of around a factor six. The flat $\alpha$ background index is estimated in the energy regions between 2.7~MeV and 3.1~MeV, and between 3.4~MeV and 3.9~MeV, excluding the $\alpha$-peak from $^{190}$Pt, originating from Pt contamination in the crucibles used to grow the crystals. In the CUORE-0 $0\nu\beta\beta$ signal region the background analysis is blinded. The blinding procedure is a form of \textit{data salting} in which a blinded portion of background events within $\pm 10$~keV of the 2615~keV peak is moved to the $0\nu\beta\beta$ Q-value, and vice versa. In this way an artificial \textit{salted peak} is created at the Q-value (figure~\ref{fig:salted}, left). The peak at the $0\nu\beta\beta$ Q-value is an artifact used to blind the energy spectrum in the signal region. The peak at 2505~keV is produced by the sum energy of the two $\gamma$-rays originating from $^{60}$Co contamination in the copper.

The flat background in the ROI is (0.063$\pm$0.006)$\,$counts/(keV$\cdot$kg$\cdot$y) while it was (0.153$\pm$0.006)$\,$counts/(keV$\cdot$kg$\cdot$y) in Cuoricino, a reduction around a factor 2.5. The energy resolution in the ROI is defined as the FWHM of the 2615~keV $\gamma$-ray peak, and a fit to the summed background spectrum of all fully functional channels gives a value of 5.1~keV. With this resolution and this background index, the CUORE-0 sensitivity will overcome the Cuoricino half life limit (which is $2.8\times 10^{24}$~y at 90\% C.L.~\cite{CUORICINO}) within about one year of live time (figure~\ref{fig:salted}, right). 
 
\section{Conclusions}
The bolometric technique is one of the most promising approaches for rare event searches. After the successful operation of Cuoricino, the CUORE collaboration is building a 741$\,$kg array of TeO$_2$ bolometers to search for neutrinoless double beta decay in $^{130}$Te. The detector is being constructed at LNGS and the cool down is scheduled to take place in 2015. The potential of CUORE is demonstrated by the pilot CUORE-0 experiment, a 39$\,$kg array of TeO$_2$ bolometers that has been operating at LNGS since March 2013. The CUORE-0 detectors showed a FWHM energy resolution of 5.1~keV at 2.6$\,$MeV, and a flat background of (0.063$\pm$0.006)$\,$counts/(keV$\cdot$kg$\cdot$y) in the signal region. The background above 2.6$\,$MeV, originating from $\alpha$-decays occurring on the surfaces of the crystals and of the copper constituting the tower structure, was measured to be (0.020$\pm$0.001)$\,$counts/(keV$\cdot$kg$\cdot$y), a factor of six better than in Cuoricino. The background above 2.6$\,$MeV gives an estimate of what is expected to be the main background contribution in the signal region in CUORE.
The combination of the CUORE-0 background measurement and Monte Carlo simulation of the CUORE geometry, predicts that after five years of live time CUORE will reach a 90\% C.L. sensitivity of 9.5$\times10^{25}\,$y on the $0\nu\beta\beta$ half life of $^{130}$Te.

\section*{Acknowledgments}
The CUORE Collaboration thanks the directors and staffs of the Laboratori Nazionali del Gran Sasso and the technical staff of our laboratories. This work was supported by the Istituto Nazionale di Fisica Nucleare (INFN); the National Science Foundation under Grant Nos. NSF-PHY-0605119, NSF-PHY-0500337, NSF-PHY-0855314, NSF-PHY-0902171, and NSF-PHY-0969852; the Alfred P. Sloan Foundation; the University of Wisconsin Foundation; and Yale University. This material is also based upon work supported by the US Department of Energy (DOE) Office of Science under Contract Nos. DE-AC02-05CH11231 and DE-AC52-07NA27344; and by the DOE Office of Science, Office of Nuclear Physics under Contract Nos. DE-FG02-08ER41551 and DEFG03-00ER41138. This research used resources of the National Energy Research Scientific Computing Center (NERSC).

%

\bibliography{GIACHERO_Andrea_ICNFP2014}

\end{document}